# Drift-insensitive distributed calibration of probe microscope scanner in nanometer range: Real mode

Rostislav V. Lapshin[1, 2]

[1]*Solid Nanotechnology Laboratory, Institute of Physical Problems, Zelenograd, Moscow, 124460, Russian Federation*

[2]*Moscow Institute of Electronic Technology, Zelenograd, Moscow, 124498, Russian Federation*

E-mail: rlapshin@gmail.com

A method is described intended for distributed calibration of a probe microscope scanner consisting in a search for a net of local calibration coefficients (LCCs) in the process of automatic measurement of a standard surface, whereby each point of the movement space of the scanner can be defined by a unique set of scale factors. Feature-oriented scanning (FOS) methodology is used to implement the distributed calibration, which permits to exclude *in situ* the negative influence of thermal drift, creep and hysteresis on the obtained results. The sensitivity of LCCs to errors in determination of position coordinates of surface features forming the local calibration structure (LCS) is eliminated by performing multiple repeated measurements followed by building regression surfaces. There are no principle restrictions on the number of repeated LCS measurements. Possessing the calibration database enables correcting in one procedure all the spatial distortions caused by nonlinearity, nonorthogonality and spurious crosstalk couplings of the microscope scanner piezomanipulators. To provide high precision of spatial measurements in nanometer range, the calibration is carried out using natural standards – constants of crystal lattice. The method allows for automatic characterization of crystal surfaces. The method may be used with any kind of scanning probe microscope (SPM).

## 1. Introduction

Usually, a probe microscope scanner is characterized by three calibration coefficients $K_x$, $K_y$, $K_z$ representing sensitivities of X, Y, Z piezomanipulators, respectively (to take into consideration a possible nonorthogonality of X, Y piezomanipulators, an obliquity angle should be additionally determined) [1, 2, 3, 4]. Because of piezomanipulator's nonlinearity [5, 6] and spurious crosstalk couplings, the probe microscope scanner may be described by the above coefficients only near the origin of coordinates, where the influence of the distortion factors is insignificant. As moving away from the origin of coordinates, the topography measurement error would noticeably increase reaching the utmost value at the edge of the scanner field (see Sec. 2.5).

The problem may be solved by using a distributed calibration, which implies determining three local calibration coefficients (LCCs) $K_x$, $K_y$, $K_z$ for each point of the scanner movement space, which can be thought of as scale factors for axes *x*, *y* and *z*, respectively [7, 8, 9, 10, 11, 12]. A reference surface used for calibration should consist of elements, called hereinafter features, such that the distances between them or their sizes are known with a high precision. The corrected coordinate of a point on the distorted image of an unknown surface is obtained by summing up the LCCs related to the points of the movement trajectory of the scanner [10].

# Drift-insensitive distributed calibration of probe microscope scanner

Both lumped and distributed calibration of the probe microscope scanner should be carried out by the data where distortions caused by drifts (thermal drifts of instrument components plus creeps of piezomanipulators) are eliminated. Otherwise, the measurements will have large errors [2, 3, 11, 13, 14]. In the present work, to eliminate the negative influence of thermal drift and creep on the distributed calibration results, the methods are used of feature-oriented scanning (FOS) and of counter-scanning suggested in Refs. 10, 13, 14.

## 2. Experimental results

Distributed calibration of scanner of the probe microscope Solver™ P4 (NT-MDT Co., Russia) in the real mode was carried out by atomic topography of highly oriented pyrolytic graphite (HOPG) monocrystal basal plane (0001). In order to minimize thermal deformation of the sample, a graphite crystal of small dimensions 2×4×0.3 mm was used. Three adjacent carbon atoms (or interstices) forming an equilateral triangle ABC were selected as a local calibration structure (LCS) [10]. According to the neutron diffraction method, the HOPG lattice constant $a$ (i. e., side length of the ABC triangle) makes 2.464±0.002 Å [15].

Graphite surface topography was obtained by scanning tunneling microscopy (STM). With microscope Solver, scanning is carried out by moving a sample relative to a fixed tip. As the tip, a mechanically cut ∅0.3 mm NiCr wire was used. To protect the microscope against seismic oscillations, a passive vibration isolation system was employed. Moreover, the microscope was housed under a thermoinsulation hood, which also served as an absorber of external acoustic disturbances. Neither constant temperature nor constant humidity were maintained in the laboratory room where the calibration was implemented. The typical noise level of the tunneling current in the course of the measurements made about 20 pA (peak-to-peak).

The microscope scanner is a piezotube having four electrodes on the outer side and a solid grounded electrode inside. The piezotube was fabricated by sintering piezoceramic powder mixture similar to PZT-5A (lead zirconate titanate Pb(Zr, Ti)$O_3$, ELPA Co., Russia). The maximal unipolar voltage applied to the piezotube X, Y electrodes makes 150 V. Between the electrodes of the manipulators of the same name, the tube piezoceramic is polarized in mutually opposite directions. X, Y DACs have capacity of 16 bits.

During the raster scanning, the probe movement velocity at the retrace sweep was set the same as at the forward trace. Immediately before every raster scanning, a "training" of the scanner was carried out [14]. The scanner training is a repeated movement along the first line, which allows to decrease creep at the beginning of the scan [16]. While training, the actual scanning velocity was also determined.

*2.1. Selection of a calibration area on the surface of a standard*

Since the whole scanning field of the microscope used makes about 2×2 μm and the typical sizes of atomically smooth regions suitable for calibration on the surface of the graphite sample did not usually exceed 450×450 nm, the distributed calibration was carried out by parts. First, a trial scanning of the whole field was conducted, and then plane smooth typically 300×300 nm size regions with no visible defects were selected therein.

The above scale is set based on the following considerations. First, nanodimensional surface defects, whose presence is undesirable, are still distinguishable with this scale. Second, with the typical net step of 5-25 nm, the time of calibration by a region of such sizes will not be too long. Third, around the region, a surface area suitable for calibration can still be reserved with the sizes that allow the probe to stay within its limits even after being shifted by the end of calibration due to drift.

After the calibration had been completed at each of the selected regions, the whole scanner field was shifted to

R. V. Lapshin

an adjacent area on the crystal or the pyrographite crystal was cleft. Periodical graphite cleaving not only permits to refresh the surface but also to reduce the errors resulted from small variations of the lattice constant and variations of angles between the crystallographic directions during the subsequent averaging. Those errors arise from imperfections relating to the standard itself (see Sec. 2.6).

Since during the time of calibration, which may take up from several hours to several days, the standard surface is shifted significantly within the field of view of the microscope because of drift, the whole scanner field does not require to be shifted frequently. The calibration can be continued at either the shifted old surface region or at a new region that appeared in the field of view of the instrument. The operations described above are repeated until most of the scanner field is calibrated (see Fig. 1).

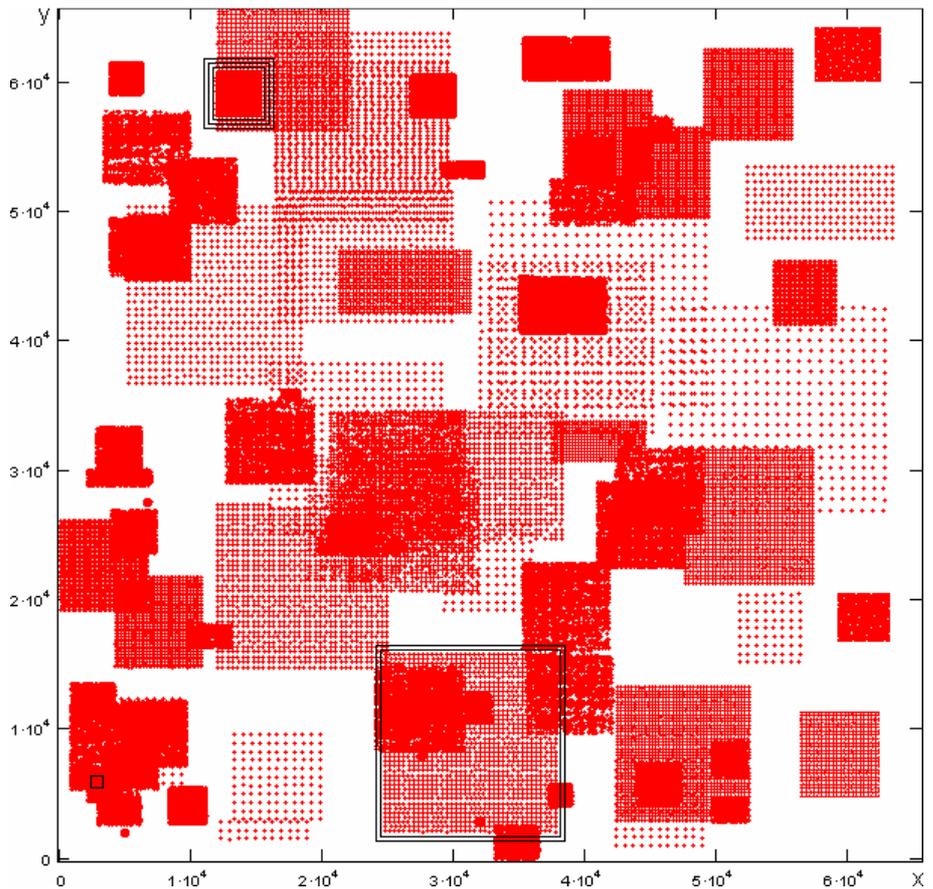

Fig. 1. Areas of distributed calibrations (squares and rectangles). The areas consist of points corresponding to LCS positions for which LCCs were found. The sizes of the areas are determined by the net step and the number of net nodes. Some areas are superimposed one on another. Outlined with a single frame is the area 73×73 Å calibrated by net of 31×32 nodes having the step as small as 8 positions (≈2.44 Å). Outlined with a double frame is the area 396×402 nm calibrated by net of 37×39 nodes having the step as large as 361 positions (≈11 nm). Outlined with a triple frame is one of the areas where variations of constant lattice of the used HOPG crystal were analyzed. The $x$ and $y$ axes are graduated in lateral positions of the scanner. The whole scanner field is approximately equal to 2×2 μm. The calibrations have been conducted at the position of the scanner Z manipulator extended to the middle of its range.

One placement of the STM head and one tip were sufficient for 1-1.5 week of ceaseless operation. The time mentioned is limited by the tip lifetime which is determined by the slow growth of a native oxide layer [17, 18]. The tip surface oxidation increases the noise level of the tunneling current which, at a certain moment, leads to the loss of atomic resolution. Graphite cleavage was carried out rarely, approximately once in 2-3 months. In case of failure to get a stable image of good quality by using the constant-current mode, the constant-height mode was used for calibration in the lateral plane.

*2.2. Measurement parameters and modes*

Before the calibration started, the scanner's Z manipulator was moved to the middle of its range by using a coarse approach stage whereupon it was held near this position during all the time of calibration. Since the drift in the vertical plane is tens of times less than the drift in the lateral plane, control commands were applied to the Z manipulator very rarely.

# Drift-insensitive distributed calibration of probe microscope scanner

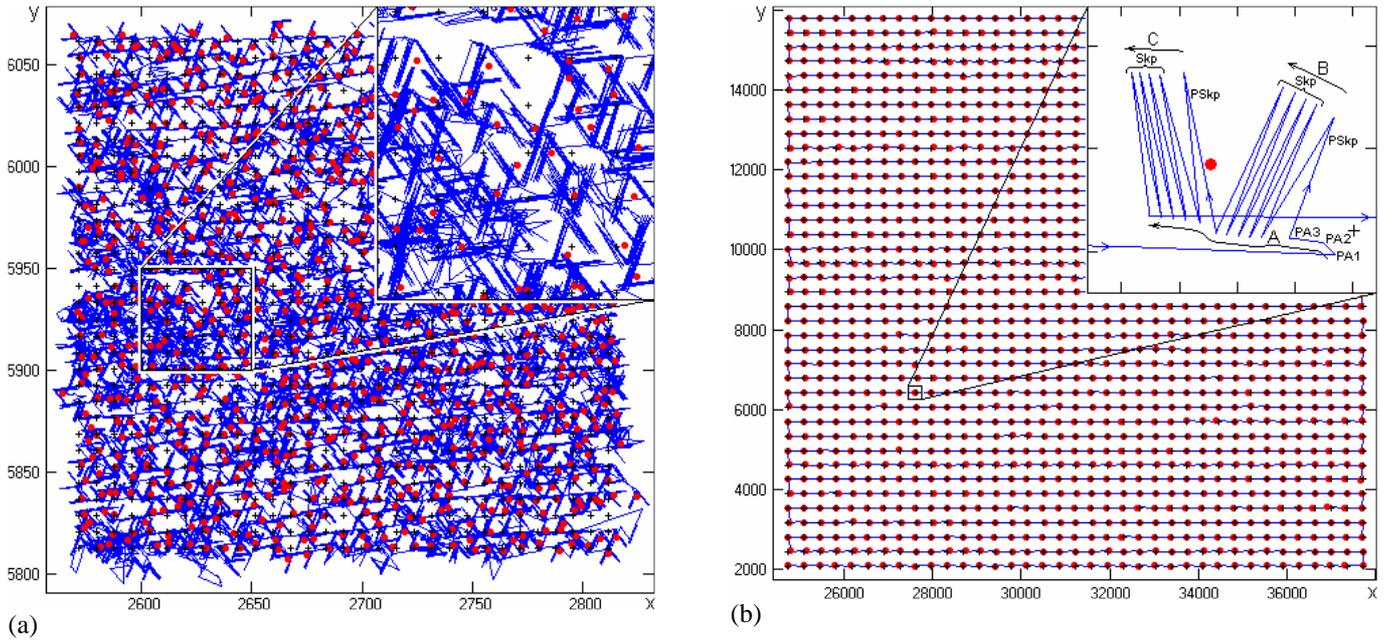

Fig. 2. LCS positions (designated as "●") for which piezoscanner LCCs and local obliquity angles were found during the distributed calibration. Calibration with (a) the small step of 8 positions (≈2.44 Å), (b) the large step of 361 positions (≈11 nm). Nodes of the initial net (a) 31×32, (b) 37×39 are shown as "+". Size of the area of graphite surface (a) 73×73 Å (the area in Fig. 1 is outlined with a single frame), (b) 396×402 nm (the area in Fig. 1 is outlined with a double frame). Trajectory of probe movement from feature to feature is imaged with a solid line. In the inset to figure (b): PA1, PA2, PA3 are three consecutive probe attachments to the feature A; PSkp, Skp are pre-skipping and skipping cycles, respectively; the arrows near the A, B, and C features show the drift direction. Measurement mode: STM, constant-current, $U_{tun}$=50 mV, $I_{tun}$=856 pA. Waiting time after movement to the next net node (a) 3 s, (b) 11 s. Scanned apertures (a) 2429, (b) 3010. Scanned segments (a) 22750, (b) 31569. Carried out attachments (a) 24174, (b) 48365. Number of detected LCSs (a) 992, (b) 1443. Fixed feature losses (a) 17, (b) 68. Mean lateral drift velocity $\left|\bar{v}_{xy}\right|$ (thermal drift + creep) (a) 0.147 Å/s, (b) 0.149 Å/s. Calibration time (a) 23.8 h, (b) 57.3 h.

The STM measurements of the standard surface were conducted under ambient conditions at room temperature after the instrument had been warmed up [19, 20] in active state for hours [13]. The active state implies periodical attachment of the microscope probe to the carbon atom [10] located near the start position of the distributed calibration. Beside warming up, the attachment allows for monitoring the instrument state – noise level, resolution, and drift velocity. Thus, by observing the change in drift velocity in this mode, the termination moment of the warming-up stage can be determined.

The calibration results for each area were put into the common calibration database (CDB). The calibration was carried out either by LCS consisting of carbon atoms or by LCS consisting of interstices [11] depending on which LCS was closer to the current net node. Such approach allows avoiding calibration interruption in case the image of the atomic surface becomes accidentally inverted. Moreover, the probe path from the current net node into the A feature position becomes shorter.

Fig. 2 shows the initial net, the LCS positions detected during calibration as well as a movement trajectory of the probe from feature to feature (the movements while scanning apertures and segments are not shown). In Fig. 2(a), a case of calibration with the small step ≈2.44 Å is presented (net size is 31×32 nodes) and in Fig. 2(b), a case of calibration with the large step ≈11 nm is presented (net size is 37×39 nodes). In both cases, the measurements were made with the constant-current mode. According to the trajectory given in Fig. 2(a), the microscope probe is continually shifted aside from the nodes of the initial net due to some drift. Fig. 2(b) clearly shows that the movements by the net nodes are performed in full compliance with the trajectory shown in Fig. 1(a) of Ref. 10.

**R. V. Lapshin**

Only one of the LCSs detected in the aperture, the one nearest to the current net node, was used for calibration (LCS ABC, see Fig. 3(a) in Ref. 10). The aperture size was 31×31 points (9.2×8.8 Å approx.), the segment size was 21×21 points (6.1×5.9 Å approx.). The number of averagings in a raster point was 15 and 25 for calibrations in Fig. 2(a) and Fig. 2(b), respectively. The number of consecutive skipping cycles was assigned as 4 (see Skp in the inset of Fig. 2(b)). Before the skipping started, at least one preskipping cycle was implemented (see PSkp in the inset of Fig. 2(b)) [13]. The required number of preskipping cycles is assigned automatically. Preskipping is an idle (tuning) skipping required for the relative distance measurement process to reach its operating condition (see inset in Fig. 2(b), cp. trajectories of PSkp and Skp).

The scanning velocity in aperture/segment was set to 592 Å/s for calibration in Fig. 2(a) and to 465 Å/s for calibration in Fig. 2(b). Movement velocity between features was set to 22 Å/s in both cases. The number of probe attachments that follow the pause (see as an example three consecutive probe attachments PA1, PA2, PA3 to the feature A located near one of the net nodes shown in the inset of Fig. 2(b)) usually made 2-8 with the regular scanning and 2-5 with the counter-scanning. LCS measurements were performed at the moments when the lateral drift velocity did not exceed 0.05 Å/s by module.

While conducting the measurements, it was found that the larger is the speed of probe lateral movement relative to the surface in aperture/segment or at the skipping, the stronger is the creep excited during this movement. The strong creep, in its turn, requires setting a longer pause and leads to a greater number of attachments inserted automatically after the pause. On the other hand, the movement velocity should not be set too small since too slow movement leads to low calibration productivity and to higher probability of feature loss in case of a strong drift. The above scan and skipping velocities have been found experimentally and are optimal for this particular microscope.

The larger the step of the initial net, the stronger disturbance is introduced into the calibration process at the moment of movement from the current node to the next one and the longer time is required for this disturbance relaxation. The method retained full functionality while conducting distributed calibrations with the step of the initial net up to 500 nm. Larger steps were not tested because of lack of suitable regions of large enough sizes on the standard surface.

*2.3. Actions of the algorithm when the measurand goes out of the tolerance limits*

Because of defects and mostly because of all kinds of instabilities, the measured LCS sizes may not fall into the acceptance limits defined before the calibration (usually ±8% for linear dimensions and ±5° for angular ones). In such situation, the calibration program, in accordance with the operator's choice, can act on the following two scenarios. On the first one, after all the possibilities of rescans have been exhausted, the algorithm fixes a surface defect in this place, and after that it continues calibration in the next net node. In this case, LCCs for some scanner positions will be absent in the obtained CDB.

According to the second scenario, the algorithm moves the probe again into the current net node where a new attempt is made to find LCCs within the neighborhood of this node. The number of such returns to the current node depends on particular measurement conditions at the given moment of time. This state of calibration process is represented as loops on the trajectory. After some time, either the measurement conditions change so allowing to obtain LCS with parameters within the limits of the set tolerances, or, due to drift, some other area of the standard surface with no defects gets into the current node surroundings. The results shown in Fig. 1 were mainly obtained

# Drift-insensitive distributed calibration of probe microscope scanner

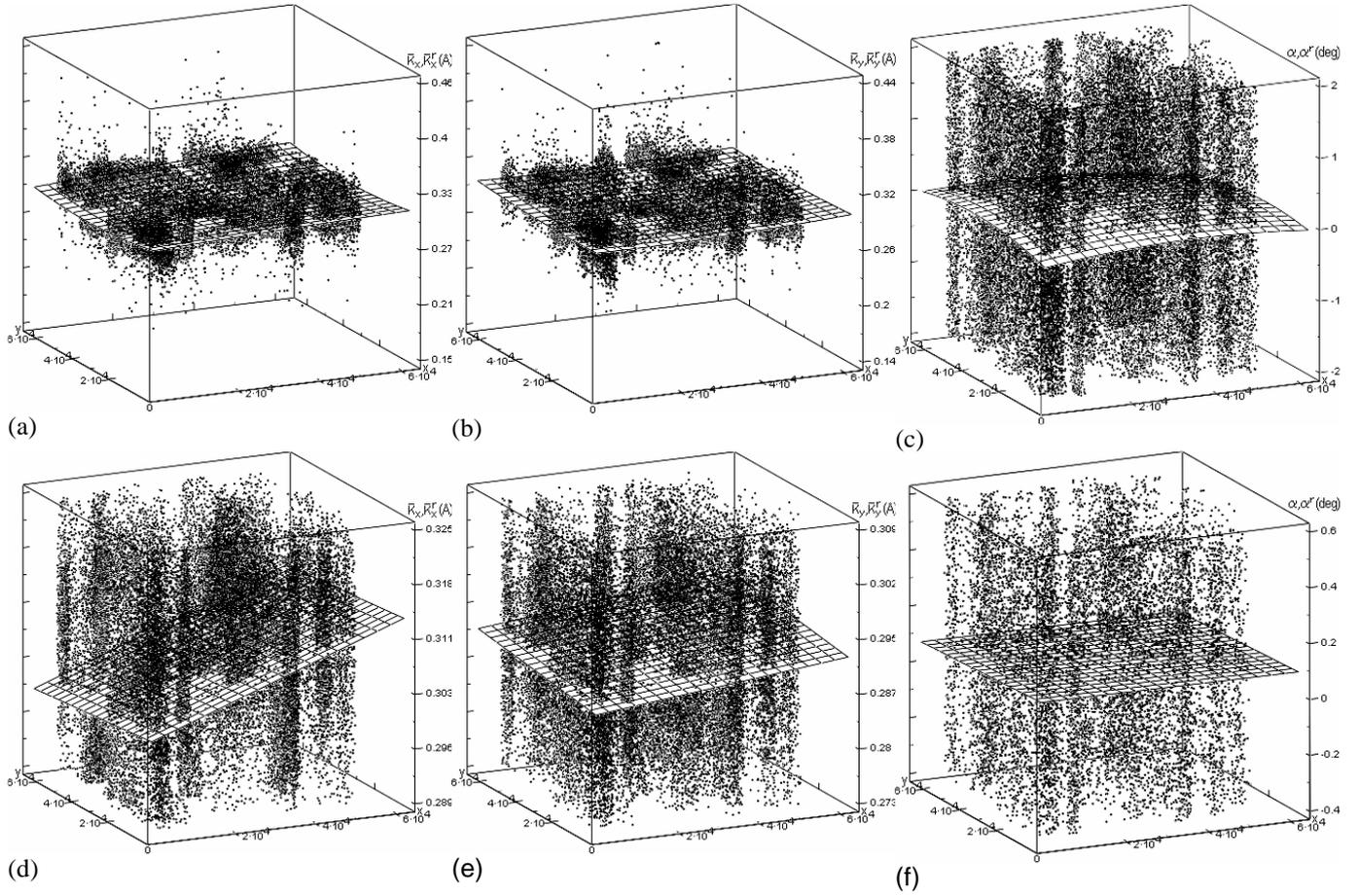

(a) (b) (c)
(d) (e) (f)

Fig. 3. Regression surfaces of 2nd order (upper row) and 1st order (lower row) drawn through (a), (d) LCCs $\overline{K}_x$, (b), (e) LCCs $\overline{K}_y$, and (c), (f) local obliquity angles $\alpha$ of the obtained CDB. The vertical scale of the surfaces (d)-(f) is intentionally stretched out so to better see the slope of those surfaces. CDB size makes 106683 LCSs. The data correspond to the position of scanner Z manipulator moved to the middle of its range.

with the use of the second scenario.

*2.4. Calibration database, constructing regression surfaces*

The sought for 2nd order regression surfaces built by LCCs of the joint CDB of the piezoscanner are presented in Fig. 3(a)-(c). The joint CDB contains information about $n$=106683 LCSs. Because of disturbing factors and, to a much lesser extent, defects of the standard, some LCCs and local obliquity angles have noticeable deviations (within the limits of the tolerances set at the calibrations) from the expected values. Therefore, before being used the CDB was thinned out discarding such values. In order to perform the thinning as efficiently as possible, first the expected values should be obtained as close to the true values as possible. To do so, the method described in Ref. 14 can be employed for example.

Mean values of LCCs $<\overline{K}_x>$=0.304 Å, $<\overline{K}_y>$=0.290 Å and obliquity angle $<\alpha>$=0.1° were found near the scanner's origin of coordinates (area of analysis 500×500 nm) using the regression surfaces. Indication that the real mode calibration is done correctly, is the proximity of the mean values of LCCs to each other [14] and to the initial coefficients $\Delta_x$=0.306 Å, $\Delta_y$=0.307 Å as well as the proximity of the mean obliquity angle to $\alpha$=0.4° [14]. Note that the coefficients $\Delta_x$, $\Delta_y$ and the obliquity angle $\alpha$ were obtained near the scanner's origin of coordinates during calibration by the image which distortions induced by thermal drift and creep had been eliminated [14].

Because of the appreciable difference between $<\overline{K}_x>$ and $<\overline{K}_y>$ (4.6% instead of 0.3% in Ref. 14), it could

**R. V. Lapshin**

be assumed that for some reason the *y* drift component was not fully corrected in the course of the distributed calibration. However, after switching the direction of the fast scan in apertures and in segments from *x* to *y*, exactly the same values were obtained. In order to finally eliminate any doubts concerning the differences in the coefficients, the sample was rotated by 90°. Had the drift been completely ruled out in the course of the distributed calibration and had the used standard have just insignificant uncertainties, the same coefficients would have been obtained over again with the new position of the sample. And indeed, after the calibration, the coefficients have not changed.

Neither has dependence been detected of the difference in the coefficients upon the step sizes of the net. It is also worth noting that the difference of the coefficients is not only observed in the neighborhood of the origin of coordinates but in other parts of the scanner field and for the whole scanner as well. For instance, the mean values calculated by the whole CDB are as follows: $<\overline{K}_x>=0.306$ Å, $<\overline{K}_y>=0.292$ Å (difference 4.6%), $<\alpha>=0.1°$. Thus, the obtained result shows that the difference between the coefficients is caused by some systematic error which has nothing to do with the standard nor with the method used for compensation of drift influence.

Since the counter-scanned images used for calibration in Ref. 14 were obtained more than 8 years ago (both then and now the same pyrographite crystal is used), it is quite likely that during such a long period of time, a certain degradation of piezoceramics, mechanical or electronic units of the microscope may have occurred. Taking into account that $<\overline{K}_x>$ exactly coincided with $\Delta_x$, it is reasonable to assume that the piezoceramics itself has not changed, since it is hardly possible that exactly half of the scanner has aged while the other one has not. Another fact against the assumption of unfavorable changes in piezoceramics is the following. It is Y manipulator whose sensitivity has decreased and it is well known that the load of Y manipulator during the raster scanning is much less than the load of X manipulator. Thus, we most likely observe a disadjustment of electronics of the piezoscanner Y channel that apparently happened when the instrument was being transported.

If the fact is established that the microscope scanner can be characterized by 1st order regression surfaces (see Fig. 3, see also Ref. 11), then the subsequent distributed calibrations of this or similar scanners can be substantially simplified. As is known, the position of a plane in space is uniquely defined by coordinates of its three points. Thus, by conducting the distributed calibration in three positions of the scanner field, located as far as possible from each other, the searched for regression planes can be found. To reach the required precision of correction of nonlinearities and spurious couplings, the appropriate number of repetitive calibrations should be carried out in each of the three positions. Since LCCs change very smoothly and the movement range of Z manipulator is usually by an order less than the movement ranges of X, Y manipulators, the search for LCCs can be carried out for, say, five approximately equidistant positions of Z manipulator. Thus, the distributed calibration in the space of scanner movements in that case can be reduced to measurement of LCCs in only 15 points.

*2.5. Nonlinear piezoscanner distortions and spurious couplings between manipulators*

The regression surface $\overline{K}_x^r$ (see Fig. 3(d)) is tilted towards the axis *x*, which points out a nonlinear response of the scanner X manipulator to the driving voltage applied. Besides, the regression surface is also slightly tilted towards the axis *y* indicating small nonlinear spurious couplings from other scanner manipulators, in particular, the Y manipulator. As is seen from the figure, the response nonlinearity of the X manipulator noticeably exceeds the nonlinearity due to spurious couplings. The regression surface $\overline{K}_y^r$ (see Fig. 3(e)) in comparison to the regression surface $\overline{K}_x^r$ has notably less overall tilt (the vertical scales in the figures (d) and (e) are the same), i. e., contribution

# Drift-insensitive distributed calibration of probe microscope scanner

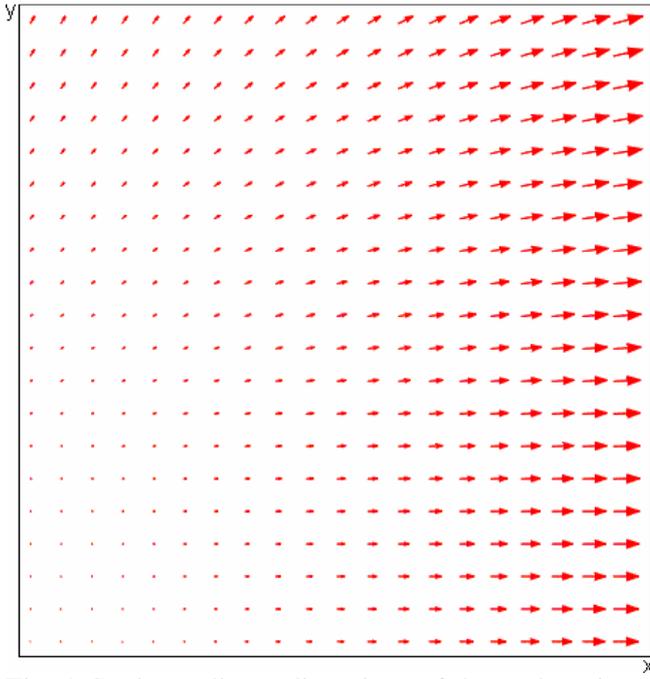

Fig. 4. Static nonlinear distortions of the probe microscope piezoscanner in the lateral plane. The arrows show the value and direction of the distortion. The length of the vector located in the right top corner of the scanner field makes 42 nm. The size of the field side is about 2 μm. The picture corresponds to the position of scanner Z manipulator moved to the middle of its range.

of Y manipulator to distortions is notably less than the one of X manipulator. It should be noted that the tilt of the regression surface $\overline{K}_y^r$ towards the axis *x* is comparable with its tilt towards the axis *y*. This means that with the Y manipulator, the nonlinearities caused by spurious couplings are comparable with piezoceramics nonlinear response. The probable cause of the observed differences between $\overline{K}_x^r$ and $\overline{K}_y^r$ is mechanical and/or material unsymmetry of the particular piezoscanner. According to Fig. 3(f), the regression surface $\alpha^r$ is located practically horizontal, therefore the dependence of nonorthogonality between the scanner manipulators X and Y on the scanner movements in the lateral plane is negligible. For numerical estimates, it is convenient to use the maximum differences $\Delta_{\overline{K}_x^r}^{\max}$=0.011 Å, $\Delta_{\overline{K}_y^r}^{\max}$=0.006 Å, and $\Delta_{\alpha^r}^{\max}$=0.03° of the regression surfaces from the horizontal plane (see equations (13) in Ref. 10), i. e., the surface into which the regression surfaces would degenerate in case of no distortions.

In order to better understand the degree and the character of the microscope scanner lateral static distortions, a vector field is shown in Fig. 4, where the arrows represent local values and local directions of the revealed nonlinear distortions. The pattern in Fig. 4 corresponds to the case when the entire field is "scanned" according to the raster trajectory of the direct image. Nonlinear raster correction is carried out by formulae (8) given in Ref. 10 by using the joint CDB.

As it was expected, the distortions are negligibly small near the scanner origin of coordinates (supplying voltages are equal to zero) and they increase as the probe approaches the edge of the range, where the measurement error reaches $\Delta_{\overline{K}_x^r}^{\max} 100\%/<\overline{K}_x>$=3.6%. The obtained error value shows that the contribution of nonlinearities and spurious couplings to the total raster distortion is several times less than the contribution from thermal drift and creep which makes 8-11% (see Refs. 11, 14). It can be clearly seen in Fig. 4 that the vectors originating from points on the *y* axis are located at some angle to this axis pointing out the existence of a small nonorthogonality (<α>=0.1°) [14]. The size of the longest vector (located in the right top corner of the field) makes 42 nm.

To determine the residual lateral error of the distributed calibration method, root-mean-square deviations of LCCs are found by formulae (14) given in Ref. 10: $\sigma_{\overline{K}_x}$=0.014 Å, $\sigma_{\overline{K}_y}$=0.013 Å ($\sigma_\alpha$=1.1°). Taking into account the number *n* of LCSs in the obtained CDB, the estimated error could make $3\sigma_{\overline{K}_x}/\sqrt{n}$=±0.0001 Å (3σ).

*2.6. Errors of the standard surface, automatic characterization of a crystal surface*

It was detected in the course of the scanner calibration that graphite lattice has small variations of sizes in dif-

**R. V. Lapshin**

ferent places of the crystal surface and in the same place of the surface before and after cleavage. Those variations were numerically estimated by comparing the LCCs from CDB that corresponded to the same area of the scanner field (in Fig. 1, one of such areas is outlined with a triple frame) but had been taken either in different places of the standard or in the same place on different cleavages. The preliminary analysis has shown that variations of constant lattice on the surface of the used pyrographite crystal make ±0.3%. The obtained error is comparable with the error (±0.1%) of measuring the graphite constant lattice in bulk by neutron diffraction [15]. Thus, the method suggested enables not only for scanner calibration but also for estimating errors of the standard itself. Because of the importance of the issue of invariability of HOPG crystal lattice constant, a special study will be conducted to cover it in detail.

By saving the relative coordinates of B and C features (skippings A↔B, A↔C) found during the calibration, it is easy to obtain a distribution of crystal lattice constants and angles between crystallographic directions over the sample surface. Thus, beside calibration, the developed method can also be used for characterization of crystal surfaces, localization of surface and subsurface defects, studying superlattices [21], surface reconstructions, adsorption/desorption processes, etc. [22].

When implementing the surface characterization, instead of rigid binding of the net to the absolute coordinates of the scanner; it is necessary to calculate each next absolute net position relative to the position of the feature A determined in the end of the last skipping cycle A↔C (see inset in Fig. 2(b)). In this case, the microscope probe will drift along with the surface exactly like in FOS [13, 21, 23] and it will never leave the limits of the crystal area under characterization.

## 3. Discussion

At present, microscopes equipped with a closed-loop positional system are unable to measure surface topography with high resolution because of noises of the linear position sensors. To operate microscope near or at the utmost resolution, the closed-loop positional system has to be turned off. As a result, the measured topography is distorted by drifts, nonlinearities, and spurious couplings of the scanner manipulators.

To correct the errors caused by the manipulators' nonlinearities and spurious couplings, with the closed-loop positional system turned on, a sort of distributed scanner calibration is carried out by the standards which do not require the utmost resolution of the microscope [8]. The data obtained during such distributed calibration, which describe the nonlinear scanner behaviour, are used later to correct the topography measured with a high resolution.

The fundamental distinction of the suggested distributed calibration method is that the random velocity fluctuations of the drift thermal component have no influence on the obtained LCCs. Moreover, the error of LCC determination can be substantially reduced by performing repeated scanner calibrations the number of which has no principle restrictions. Unlike the closed-loop positional system, the developed calibration approach does not introduce any additional noise which permits calibrating the scanner by natural references such as constants of crystal lattices, using separate atoms and interstitials as features.

## 4. Conclusion

The use of the atomic surface of a crystal as a standard surface allows obtaining LCC distribution of a very high density. However, at present it is impossible to cover the whole scanner's movement range with such a dense net, especially at several levels, because of low performance of the existing microscopes. The results obtained in the course of real distributed calibration show that the dense net should not necessarily be used. For example, to

**Drift-insensitive distributed calibration of probe microscope scanner**

acquire an adequate CDB, it is enough to perform calibration only in three points of the scanner field for each of 3-5 equidistant positions of the Z manipulator.

## Acknowledgments

This work was supported by the Russian Foundation for Basic Research (project 15-08-00001) and by the Ministry of Education and Science of Russian Federation (contracts 14.429.11.0002, 14.578.21.0059). I thank O. E. Lyapin and Assoc. Prof. S. Y. Vasiliev for their critical reading of the manuscript; Dr. A. L. Gudkov, Prof. E. A. Ilyichev, Assoc. Prof. E. A. Fetisov, and late Prof. E. A. Poltoratsky for their support and stimulation.

**R. V. Lapshin**